\documentclass{iaus}
\usepackage{graphicx}

\title[Planets around  giant stars] 
{Testing planet formation theories with Giant stars}

\author[Luca Pasquini et al. ]   
{Luca Pasquini $^1$ , M.P. D\"ollinger $^1$, A. Hatzes $^2$, J. Setiawan $^3$, 
L. Girardi $^4$,  L. da Silva $^5$, J.R. de Medeiros $^6$
 \and A. Weiss $^7$}

\affiliation{$^1$ESO \\ Garching ,
Germany \\ email: {\tt lpasquin@eso.org, mdoellin@eso.org} \\[\affilskip]
$^2$ Th\"uringer Landessternwarte\\  Tautemburg,  Germany \\email: {\tt artie@tls-tautenburg.de} \\[\affilskip]
$^3$ MPiA \\  Heidelberg,  Germany \\email: {\tt setiawan@mpia-hd.mpg.de}\\[\affilskip]
$^4$ INAF-OaPD \\  Padova,  Italy \\email: {\tt leo.girardi@oapd.inaf.it}\\[\affilskip]
$^5$ ON \\  Rio de Janeiro,  Brazil  \\email: {\tt licio@on.br}\\[\affilskip]
$^6$ UFRN\\ Natal,  Brazil \\email: {\tt renan@dfte.ufrn.br}\\[\affilskip]
$^7$ MPA \\  Garching ,  Germany \\email: {\tt weiss@mpa-garching.mpg.de}}
\pubyear{2008}
\volume{249}  
\pagerange{119--126}
\jname{Exoplanets: Detection, Formation and Dynamics}
\editors{J.L. Zhou, Y.S. Sun \& S. Ferraz-Mello, eds.}
\begin{document}

\maketitle

\begin{abstract}
Planet searches around evolved giant stars  are bringing new
insights to  planet formation theories by virtue of the broader stellar mass
range of the host stars compared to the solar-type stars that have  been
the subject of most current planet searches programs.
These searches among giant stars are producing 
extremely interesting results. Contrary to 
main sequence stars planet-hosting giants do not show a tendency of being
more metal rich. Even if limited, the statistics also suggest 
a higher frequency of giant planets 
(at least 10 $\%$ ) that are more massive 
compared to solar-type main sequence stars.

The interpretation of these results is not straightforward. 
We propose that the lack of a metallicity-planet connection among giant
stars is due to pollution of the star while on the main sequence, followed
by dillution during the giant phase. We also suggest that the higher mass 
and frequency of the planets are due to the higher stellar mass.
Even if these results  do not favor a specific formation scenario, 
they suggest that planetary formation  might be more complex than what 
has been proposed so far, perhaps with two mechanisms at work and one or the other 
dominating  according to  the stellar mass.  We finally stress as the detailed study of the 
host stars and of the parent sample is essential to derive firm conclusions.
\end{abstract}

\firstsection 
\section{Introduction}

\begin{figure}[b]
\begin{center}
  \includegraphics[width=3.4in]{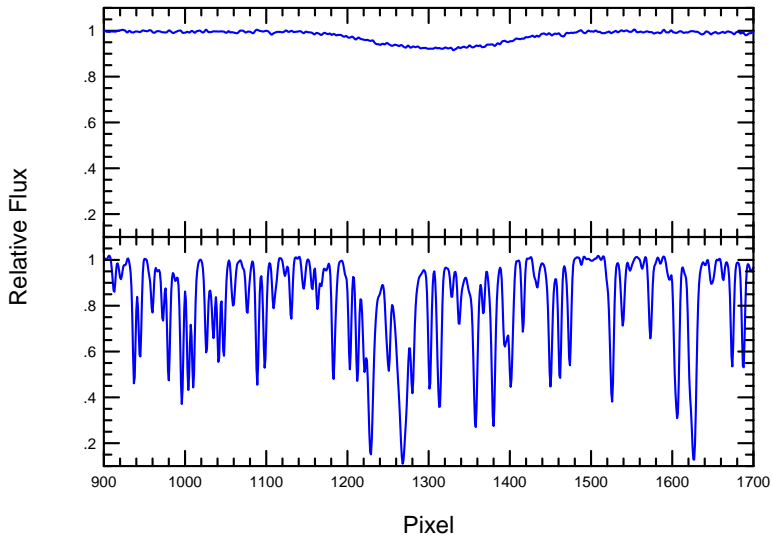} 
 \caption{Portion of spectrum in the optical region of two stars:  one massive main sequence 
 star and one cool giant. This comparison  shows that as soon as the star evolves from the main sequence migrating  in the giant domain, 
 it develops a very crowded line system, suitable for precise Radial 
 Velocity (RV) measurements. }
   \label{fig1}
\end{center}
\end{figure}

Out of the more than 200 exoplanets known, only an handful orbit around 
evolved giants stars.  These stars, however, 
are interesting targets for planet searches. Unlike the host stars of most
known exoplanets which have a mass distribution that
peaks at 0.8 $\pm$0.3 solar masses, giant
stars can have masses several times this value. The radial velocity  (RV)
method, the most succesful technique at finding exoplanets, is insentive 
to more massive, early-type stars. These hot stars 
have few spectral lines that are
often broadened by high stellar  rotation rates. The typical  RV error
for a hot early-type star might be 100 -- 1000 ms$^{-1}$, much  higher than
the $\approx$ 10 ms$^{-1}$ needed to detect sub-stellar companions.
On the  other hand, early-type main sequence stars that have  evolved off the
main sequence and have become giants have a
plethora of narrow asorption lines that are amenable to RV planet searches.
Figure \ref{fig1} clearly shows this difference, where the spectral 
region of a main sequence A star is compared to the same spectral region of
a giant star with approximately the same mass.
While in  the main sequence star 
only one broad feature  is visible, in the giants many, narrow lines are present. 
\begin{itemize}
\item The first interest in searching planets around giants is therefore to enlarge the stellar mass
range surveyed, including  higher masses. 
 \end{itemize}
  
Giants differ from main sequence stars in their structure:
they have much 
larger radii (typical of 10 R$_{\odot}$ , 
see (\cite[da Silva et al. (2006)]{dasilva06})  and  they have a much
deeper convective zone compared to solar-type  stars (cfr. discussion 
 later).   Differences in  stellar radius and depth of the convection
zone  are expected also along the main sequence, but their
variations  are small compared to the large differences 
occurring when the stars evolved into giants.
So it shall be much more evident and easy  to detect  those effects, such chemical pollution, 
 whose signatures vary    with   internal structure changes. 
 
 \begin{itemize}
 \item The second interest in searching planets around giants is therefore that, thanks to their 
 different structures,  giants provide ideal testbed to understand 
 which stellar or planetary system characteristics  depend  from some   
 stellar parameters, such as depth (mass) of the convective zone or stellar  radius, which 
 change radically when passing from the main sequence to the giant phase. 
\end{itemize}

Interestingly, giants have been early recognized as radial velocity variable, and they could be 
broadly characterized by two variability timescales:

\begin{enumerate} 
\item  Short timescales , of the order of a few days or fraction of day.  
These are  most likely due to 
pulsations or solar-type oscillations (\cite[Hatzes \& Cochran (1994)]{hatzes94},
\cite[Frandsen et al. 2002]{fra02})

\item Long timescales, of the order of one hundred or a few hundred days,  
either due to planetary companions or to stellar surface structure.
 (\cite[Hatzes \& Cochran (1993)]{ch93}).

\end{enumerate}

Because giant stars have higher intrinsic ``noise'' due to stellar oscillations,
these stars have been so far neglected by planet searches. 
Although the situation is rapidly evolving, as testified by the contributions
to this conference 
(cfr. Quirrenbach et al.,  Niedzielski et al.,  Liu, Y. et al., Masasshi et al. 
these proceedings ). 
In addition, one program is dedicated at  Lick Observatory to follow up 
evolved A stars (\cite[Johnson et al. 2007a]{johnson07a}).

\section{The FEROS and the Tautenburg Surveys}

Our group started to obtain accurate radial velocity of giant stars in 1999, 
with the thesis work of J.  Setiawan (\cite[Setiawan et al. (2003b, 2004) ]{setiawan03}) 
who used the FEROS spectrograph at the 
ESO 1.5m telescope. Seventy-six stars were observed  for four years, and the most promising
were followed up  in the last years after the spectrograph 
was moved at the 2.2 m telescope,  using time allocated to 
the Max Planck Institute for Astronomy.
 We will refer in the following to this sample as the `southern' sample. 

In 2004  M. D\"ollinger  started a survey with the 2m 
Alfred Jensch telescope in Tautenburg
 on 67 stars in the northern sky. 
 
 In both cases the selected stars are giants (class III) with accurate 
 Hipparcos parallaxes, known to better than 10$\%$. 
 For  the Tautenburg sample  we added the constraint that the stars should 
 be circumpolar so that observations could be obtained throughout the year.
Intrinsically bright and thus cool giants  were  not selected for our samples
to avoid the possible occurance of pulsating AGB stars which may have
large RV amplitude variations and the  fact that it is difficult to 
characterize their stellar parameters (cfr. next sections).

\begin{figure}[h]
\begin{center}
\includegraphics[width=3.0 in]{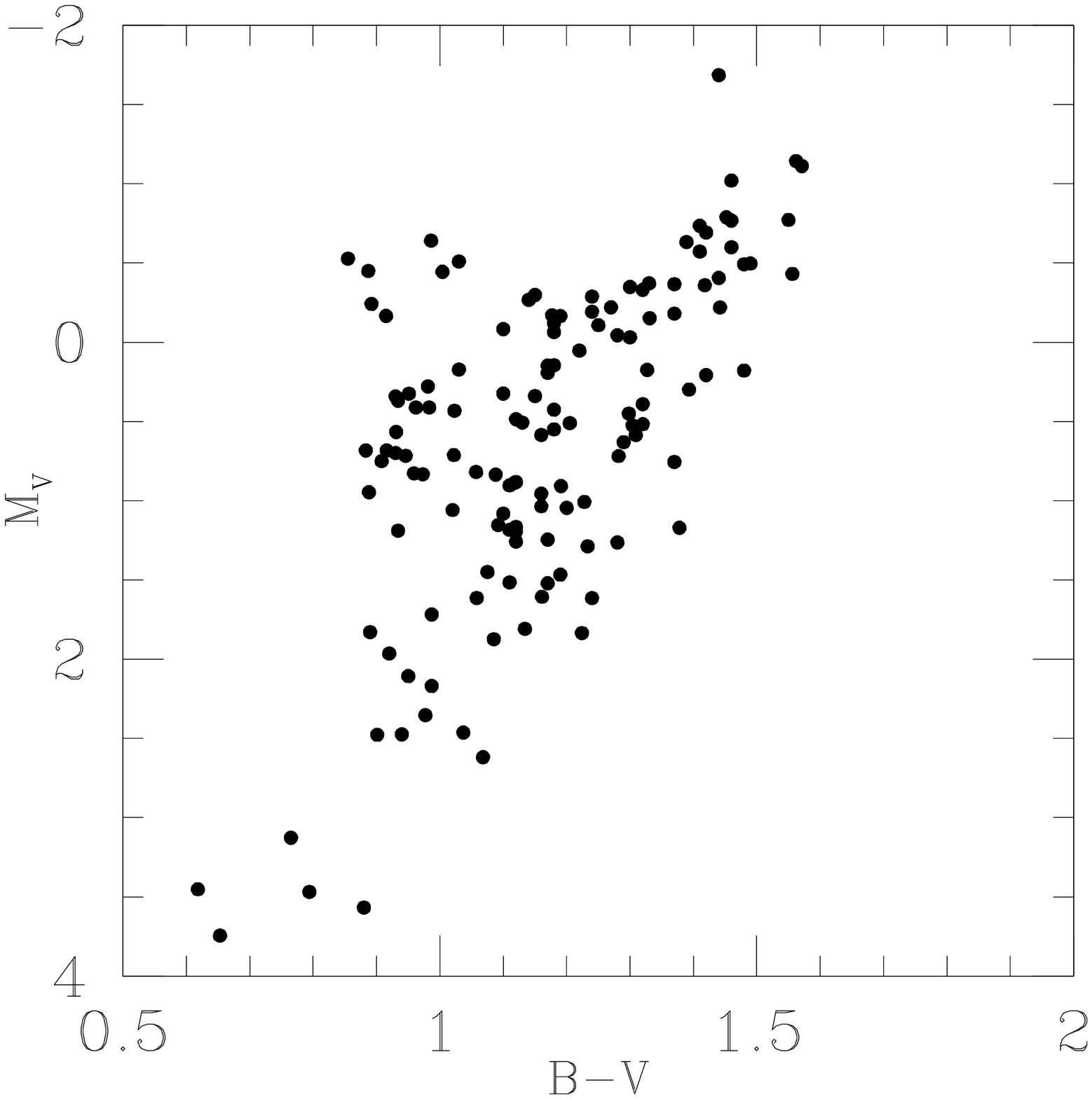} 
 \caption{Color-magnitude diagram of the  sample stars, observed with FEROS and at Tautenburg.
 The stars well sample the Red Giant Branch (RGB), the clump  and the beginning of the AGB.
 Only a few subgiants are present, and a few cool, luminous stars. }
   \label{fig2}
\end{center}
\end{figure}

 Figure \ref{fig2} shows the colour-magnitude diagram of all our  sample stars. 
 Clearly we sample well the RGB, the clump and the beginning of the AGB. 
 Very few subgiants were included.  
 
\section{Characterization of the Stars}
 
Shortly after the first exoplanet  has been 
 discovered (Mayor and Queloz 1995) it was clear 
that it is essential to characterize the host stars.
 Age, mass and metallicity are fundamental parameters to understand the  planetary
 formation mechanism and the relationship between stars and protoplanetary disks. 
 The early discovery of the 'planet-metallicity' connection
 (\cite[Gonzalez 1997]{gonzal96}, Santos et al. 2000, 2001) 
 was  exciting  and triggered 
 more observational studies and theoretical efforts. 
 When dealing with giants, the characterization of the stars is 
 particularly important and also more difficult  because  of the
presence of two degeneracies. First,
 while stars of different masses are 
 quite well separated on the main sequence, while ascending the RGB (and in the clump)
they occupy more or less  the same region of
 the color-magnitude diagram.
 Second, evolutionary tracks of metal poor giants are bluer 
 than for the tracks of metal rich giants. 
Thus an old metal poor star will occupy
 the same region of the CMD as a younger metal rich giant. 
This is the well known age-metallicity degeneracy. Thus to fully characterize
a giant star (mass, radius, etc.) using evolutionary tracks requires
 the analysis of high resolution 
 spectra from which effective temperature
 and metallicity  are derived.

 \begin{figure}[h]
\begin{center}
 \includegraphics[width=3.0in]{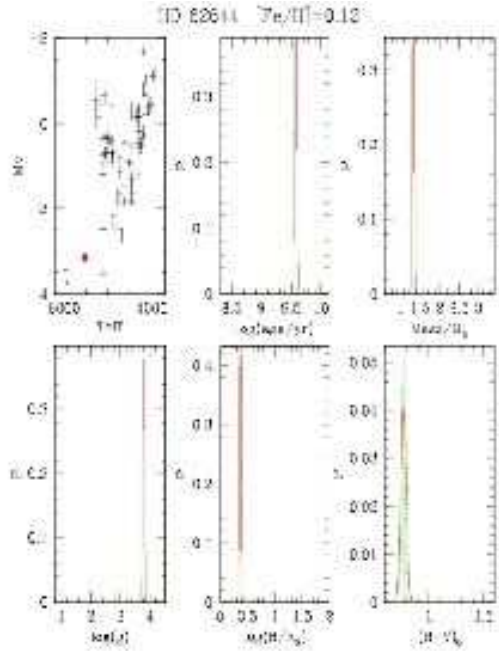} 
 \caption{ Probability Distribution Functions for 
 one star of our sample. For this object all the PDFs are very 
 well peaked and  give unambiguous results (from \cite[da Silva et al. (2006)]{dasilva06}). }
   \label{fig3}
\end{center}
\end{figure}

  Effective temperature and metallicity  were derived using 
   a classical abundance analysis in a plane parallel 
  atmospheres and  LTE  conditions. 
 From  these values  the age and mass have been derived using a modified 
version of the 
 Jorgensen and Lindegren (2005)  method   which computes the 
probability distribution function (PDF) of each 
 quantity using theoretical tracks  (cfr. \cite[da Silva et al. (2006) ]{dasilva06}).
 The exact shape of the PDF 
 differs from star to star because that depends on the 
 parameters errors and   mostly on the position  of the star in the CMD. 
In most instances the PDFs  are sharply peaked and thus result in a 
well-determined parameter
 (Figure \ref{fig3}). Some stars, however, can have 
more uncertain parameters due to a broader, or even double-peaked PDF.
 \cite[da Silva et al. (2006)]{dasilva06} have 
 performed several sanity checks on the recovered stellar parameters 
 by comparing the derived colors and  radii 
 with observed ones. They found excellent agreement  with the
values derived with our PDF method.
 
\begin{figure}[h]
\begin{center}
 \includegraphics[width=3.4in]{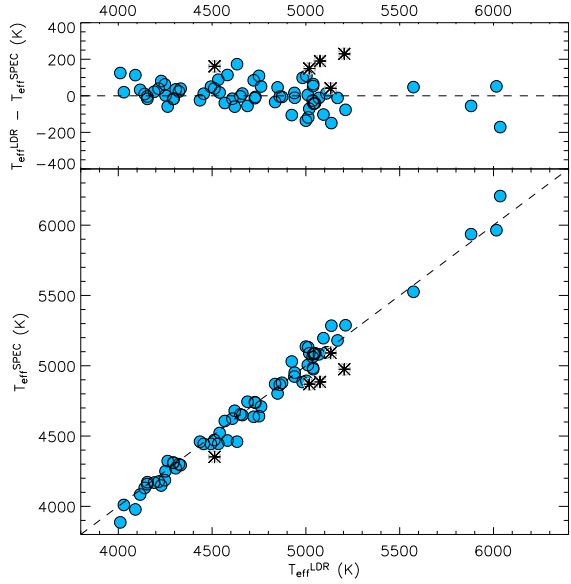} 
 \caption{Comparison between effective temperatures obtained from the abundance analysis and
 effective temperature as obtained by the Line Depth  Ratio method for the stars of the 
 southern sample. The agreement is excellent 
 (From \cite[Biazzo et al. (2007b)]{biazzo07})}
   \label{fig4}
\end{center}
\end{figure}

 We have further measured the effective temperatures of all the stars of the southern sample 
 using the line-depth ratio method (\cite[Gray 1991]{gray93}, 
 \cite[Biazzo et al. (2007a)]{biazzo07a}). They  agreed very well with the  temperatures
derived with  the abundance analysis. There was 
 a small zero point difference (5 K) and a 
 dispersion of less 70 K, as shown in Figure \ref{fig4}. 
 Finally, applying the same PDF method
to six giants of  the open cluster IC4651 we obtain for the stars 
of this cluster  a mass of 2.0
$\pm$0.2  solar mass and an age of 1.2 $\pm$0.2 Gyrs, 
very similar to what is quoted in the WEBDA database 
(\cite[Biazzo et al. 2007b]{biazzo07b}).  
 
We are  confident that we are able to produce  reliable  parameters for 
our sample stars, and we have applied the same analysis also to the 
Northern sample \cite[(D\"ollinger et al. 2008a)]{doellin08a}.

\section{Observational Results }

The number of giants hosting planets is not large enough to allow refined analysis, 
but it can  start to provide  strong indications on several aspects such as planetary frequency, 
dependence on metallicity, planets characteristics, which can be compared with 
similar properties in main sequence stars.  


\subsection{ Radial Velocity Statistics of the Surveys}

After several years of observations, we can summarize the general 
statistics we have obtained separately for the southern and northern samples.
Table \ref{tab1} summarizes our main findings in terms of RV behavior. 

\begin{figure}[h]
\begin{center}
 \includegraphics[width=3.4in]{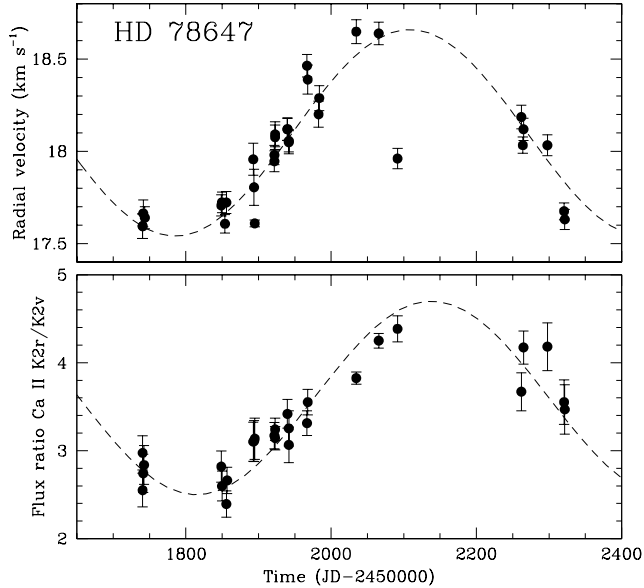} 
 \caption{Variations of chromospheric flux ( Ca II K line core ) in phase with the 
 radial velocity period of one giant. These variations indicate that the associate Radial Velocity 
 orbit is not due to the presence of planetary companion, rather by variable chromospheric activity. }
   \label{fig5}
\end{center}
\end{figure}

A few aspects of this table require some comments. 
In the introduction we have mentioned that, when analyzing giant stars, 
 special care should be taken 
in excluding other sources of RV  variability. Indeed 
out of the 32  RV variable in the Southern
hemisphere,  5 stars show associated chromospheric variability and/or bisector 
variability in phase with the RV period. An example is given in Figure \ref{fig5}, which shows 
`planet-like' RV variability that is 
correlated with Ca II core variations. Therefore  these RV variations are
likely caused by chromospheric activity rather than due 
a planetary companion. 
For the  `binaries' bin we included all stars which showed RV variability 
of several km\,s$^{-1}$ , but  for which we have insufficent data
to calculate for them a full orbital solutions. For the southern sample 
they were simply removed from future RV follow-up measurements. For the northern sample
we will continue to obtain RV measurements, but with less frequent sampling, in order to
derive the orbit.

Clearly, most numbers agree very well between the two surveys. 
The only real discrepancy is in the number of `RV constant stars', 
whose percentage is much larger in the southern sample. However, at issue
here is what is defined as a ``constant star''. The FEROS program had a 
measurement precision four times worse than for the northern sample. Not surprisingly,
this program revealed more ``constant stars''. Several of the northern sample
giant stars have RV variations of $\approx$ 20  m\,s$^{-1}$ which were clearly variable, 
but would have qualified as a constant star in the southern program. With increased measurement
precision many of the constant stars measured by FEROS may turn out to be variable.
 We therefore confirm the first conclusions obtained by e.g. \cite[Walker et al. (1989)]{wa89}
that  G and K giants are a new class of Radial Velocity variable. 

\begin{table}
  \begin{center}
  \caption{Overview of the Radial Velocity variability obtained in our surveys.}
  \label{tab1}
 { 
  \begin{tabular}{ccc}\hline 
 & {\bf Northern Sample (76)} & {\bf Southern Sample (62)} \\ \hline
 Binaries &   15 (20$\%$) & 13 (21$\%$)  \\ \hline
Variables &  32 (43$\%$)  & 22 Long Period (34$\%$) \\  
                  &  5 Activity Modulation      &  19 Short Period (31$\%$) \\ \hline
   Planets & 7 (10$\%$)  & 6 (10$\%$) \\ \hline
  Constant &  21 (27$\%$)   & 2 (3$\%$)  \\ \hline
 Precision  &  22 m/sec        &  $\sim$5 m/sec \\ \hline
  \end{tabular}
  }
 \end{center}
\vspace{1mm}
\end{table}

 {\bf The second important result is that giant planets around giant stars 
seem to be common. }
Their frequency is of at least 10$\%$, since our findings probably represent a lower limit.
We also note that the long orbital periods require observations which  extend over a baseline of 
several years. Continued monitoring 
may reveal more stars in our sample as hosting giant planets.
The planets span a range in periods between 150 and 1000 days, and in 
mass between 3 and 15 Jupiter masses. We are indeed sensitive only to 
quite massive planets, given the limited precision  of our observations and the 
fairly high intrinsic RV variability of evolved stars. {\bf  Thus 
a 10$\%$ frequency of very massive planets around more massive stars
than the sun may represent a minimum value. }
These results seem to agree well with other surveys around evolved stars, which find 
massive planets and even a good rate of brown dwarfs  orbiting around giant
stars (see i.e. \cite[Quirrenbach et al. these proceedings]{quirr08}).

\begin{figure}[h]
\begin{center}
 \includegraphics[width=3.4in]{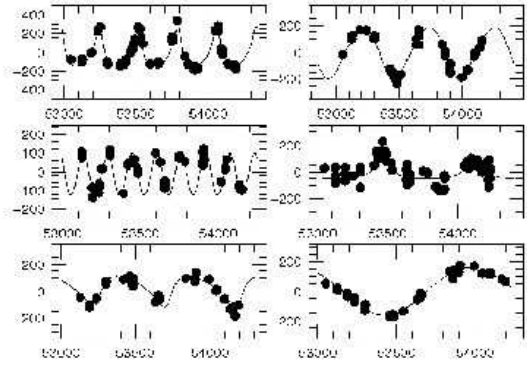} 
 \caption{Radial Velocity variations for the 6 stars of the northern sample hosting planets. 
 (D\"ollinger et al. 2008a,b in preparation)}
   \label{fig6}
\end{center}
\end{figure}

Out of the 13 exoplanets so far discovered in our surveys,  only 4  have been published
(\cite[Setiawan et al. 2003a]{set03a}, \cite[Setiawan et al. 2004]{set04}, \cite[Setiawan 
et al. 2005]{set05}, \cite[D\"ollinger et al. 2007]{doel07}), while the others are in 
preparation. We show in Figure \ref{fig6} the RV curves of the 6 planet candidates from the 
Northern sample. 

\subsection{Stellar Metallicity and Planets }

\cite[Gonzales et al. (1997, 2001)]{gon97} first  showed 
that stars hosting exoplanets tend to have
higher metallicities than stars without giant
planets.
Subsequent investigations showed this to be a real effect and not an observational bias.
(e.g. \cite[Santos et al. 2004, 2006]{santos06},  \cite[Fisher and Valenti 2005]{fv05}). 

The first determinations of abundances of planet-hosting giant stars suggested that such 
a metallicity-planet connection may not hold for evolved stars (\cite[Sadakane et al. (2005)]{sad05}
\cite[Schuler et al. (2005)]{sch05}). This was based, however,
on a small sample so no reliable conclusions could be drawn. 

\begin{figure}[h]
\begin{center}
  \includegraphics[width=3.4in]{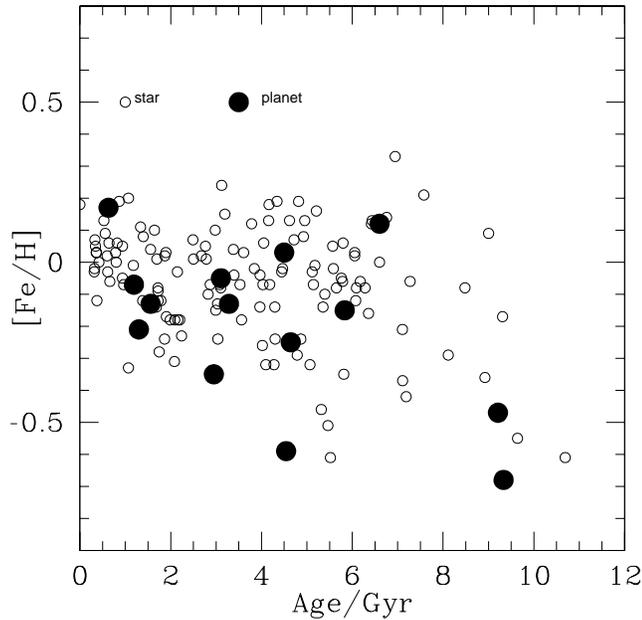} 
 \caption{Age-metallicity relationship for the giants of our Southern and Northern samples
 and for giants with planets, either from our survey or from literature. Clearly there is no difference between the two distributions. (From \cite[Pasquini et al. (2007)]{pas07}) }
   \label{fig7}
\end{center}
\end{figure}

Using the first results from our Southern and Northern surveys plus literature data 
\cite[Pasquini et al. (2007)]{pas07} have shown that this suspected trend was correct: 
giant planet-hosting stars have the same metallicity distribution as  giant stars without
planets, and have a different abundance distribution from planet-hosting  main sequence stars. 
Figure \ref{fig7} shows the age-metallicity distribution for the giants of our samples, and 
for the giants hosting planets (including data from literature), and it is clear 
that the hosting planets giants do follow the same age-metallicity relationship 
of the sample stars.

\begin{figure}[h]
\begin{center}
 \includegraphics[width=3.4in]{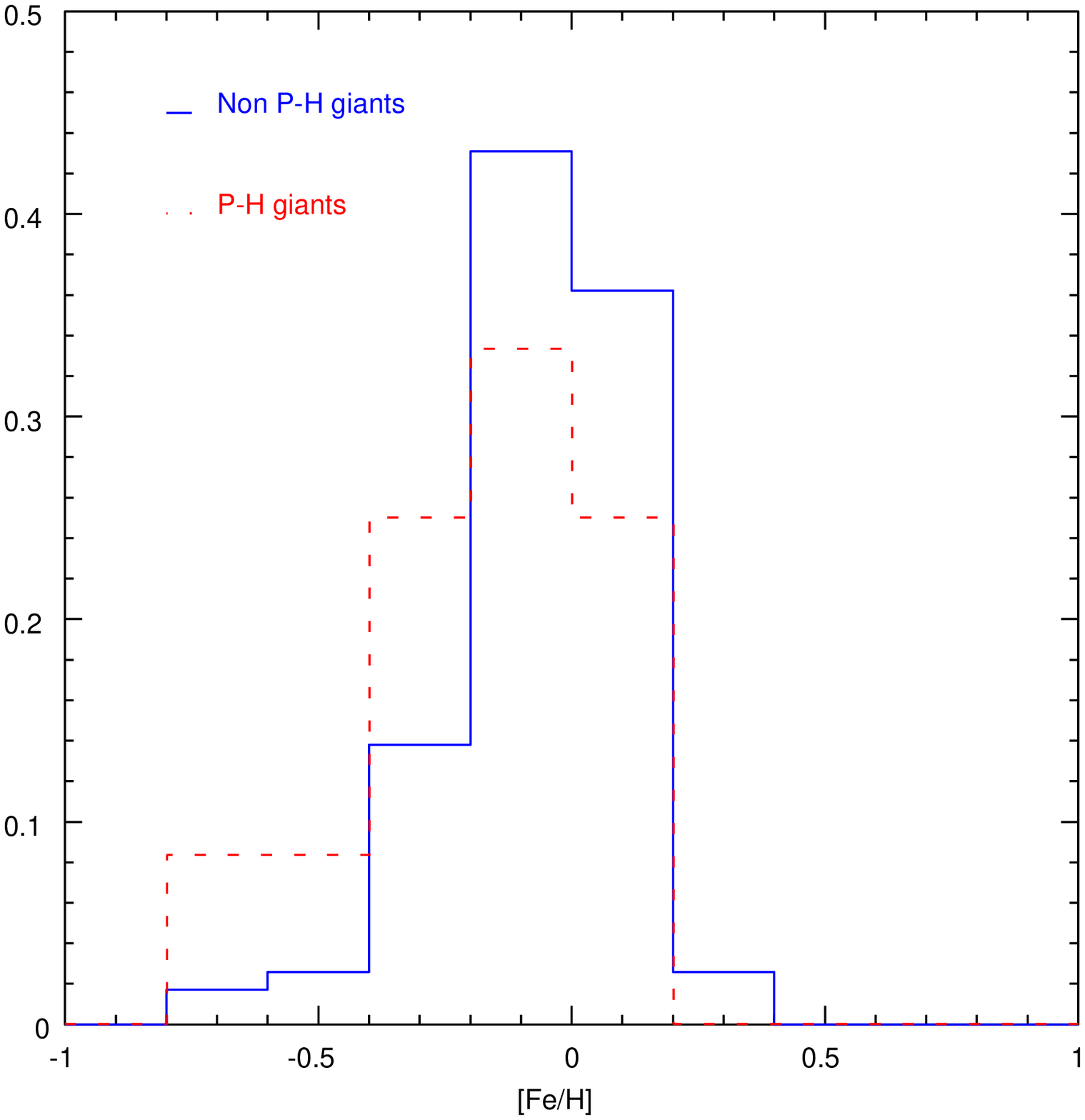} 
 \caption{Metallicity distribution of the (126) Southern and Northern giants from our 
 sample not hosting planets (Blue continuous line). With the red dashed line are indicated the 13 
 giants hosting planets, {\bf found in the same samples}. No data from literature are included. 
 This comparison clearly shows that hosting planet giants do not prefer metal rich systems.  }
   \label{fig8}
\end{center}
\end{figure}

Admittedly, the mixing of stars from literature and our own sample may introduce some bias, 
because we do not know the parent sample from which the literature data are 
taken, or  if the quoted abundances are on the same scale as the one of our sample. 
We have therefore computed the metallicity distribution of  the planet hosting (13)  
and non planet-hosting (126) stars of our samples, and this is shown in Figure \ref{fig8}. 
Clearly the two distributions overlap  and there is no indication of metal excess 
in the planet-hosting stars.

Our result are  indirectly supported by the first 5 planets included in the 
survey of evolved A stars by \cite[Johnson et al. ( 2007a)]{jon07a}, which have a 
rather uniform metal distribution. We cannot derive more firm conclusions from this sample because 
we do not know the distribution of the parent sample and 
being a young population we do expect that its age distribution should  be 
quite metal rich. 

For sake of completeness we report that  \cite[Hekker \& Melendez 2007]{hg07} 
claim a possible metal excess in their anaysis of giants hosting planets. 
We however notice two potential problems in their analysis: 
the metal excess is entirely due to the 
presence of two subgiants with exceptionally high metallicity. We have 
basically no subgiants in our sample. The second and most important
problem  is that the planet host sample used is not derived by  their parent sample: 
the planet sample is taken  from other surveys, whose metal distribution is 
not known. Clearly this is a flaw, and it is clear from their figures , which show 
that no star is present in the parent  sample in the highest metallicity bins. 
Without a knowledge of the parent sample from which the 
planet hosts stars have been derived, a  comparison is subject to 
possible bias and is thus inconclusive.

{\bf  Concerning metallicity the evidence so far collected  seems
clear: giant stars hosting planets 
do not show any tendency  towards being metal rich}. 

\subsection{Planets Characteristics}

The planets found around giant stars have long  periods (longer than 150 days), 
and relatively  large masses.   In this section we briefly investigate 
some of the planets' characteristics. 

\begin{figure}[h]
\begin{center}
  \includegraphics[width=3.0in]{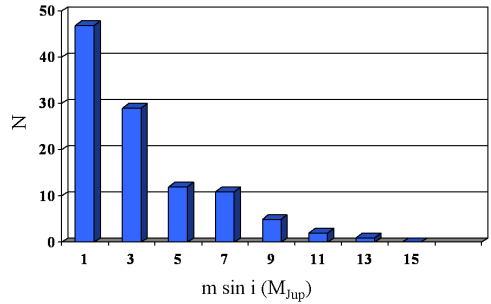} 
 \caption{Mass distribution of the planets around stars with masses below
 1.1 M$_{\odot}$ showing the strong increase for small planet mass.   }
   \label{fig8a}
\end{center}
\end{figure}

\begin{figure}[h]
\begin{center}
 \includegraphics[width=3.4in]{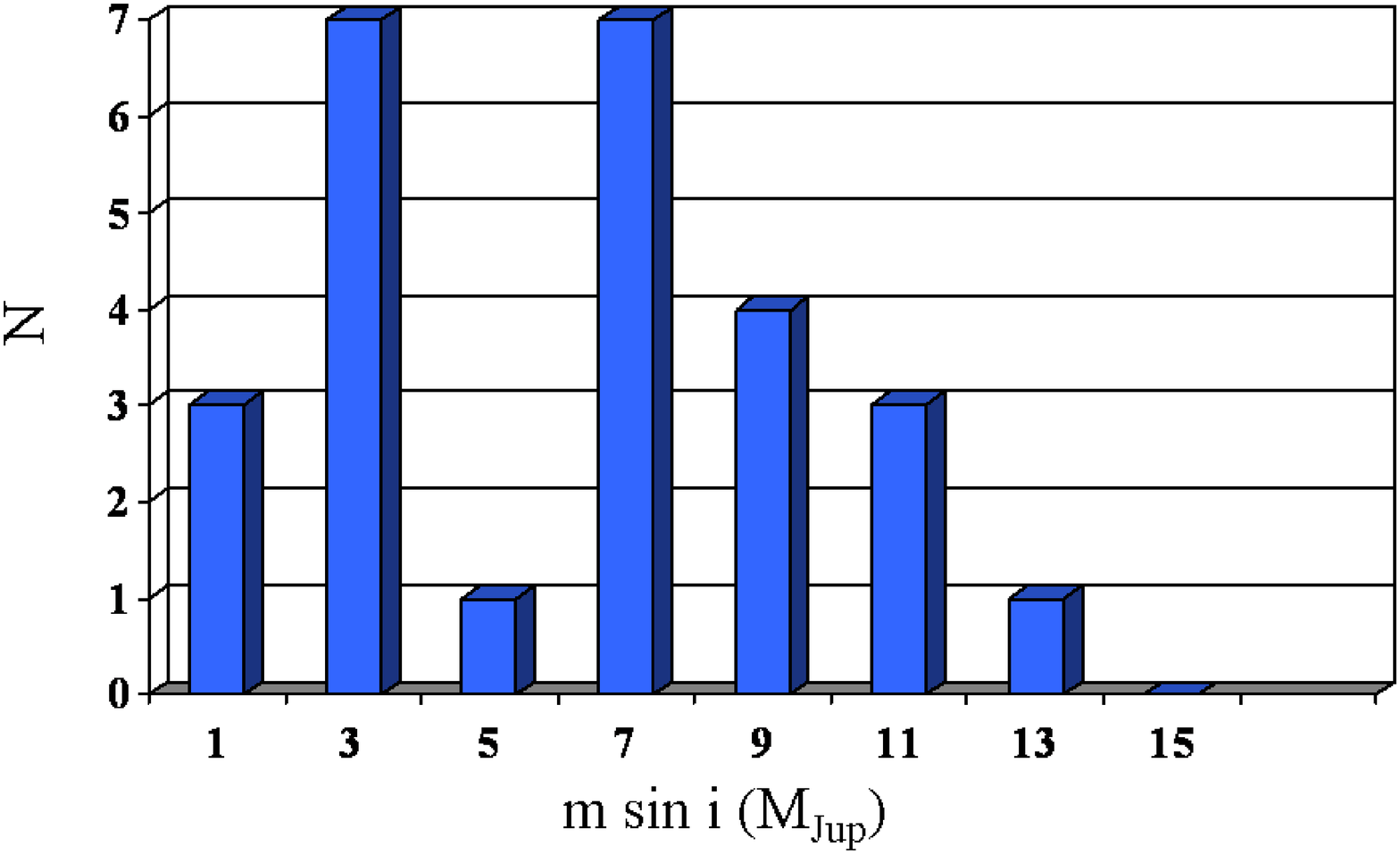} 
 \caption{Mass distribution of the planets around stars with masses above 
 1.1 M$_{\odot}$ (dominated by the giants) and for stars with smaller masses. 
 The distribution is clearly different, with  more massive stars showing a very high 
 frequency of massive planets.   }
   \label{fig8b}
\end{center}
\end{figure}

Planet masses for giant stars are typically   larger than what has
been observed around main sequence stars. 
Figure \ref{fig8a} shows the planet mass distribution for main sequence stars  with masses
below 1.1 M$_{\odot}$ while Figure \ref{fig8b} 
is the same but for stars (mostly giants) with masses larger than 1.1M$_{\odot}$.
Clearly the distribution of planet masses for low mass  main sequence stars 
increases towards lower mass planets, and  it is 
definitely  different from the one of more massive stars. 
Of course we know that there are biases in the giants' sample, in that 
small mass planets are presently out of reach of surveys due to their 
limited precision and to the intrinsic variability of the stars. Nevertheless, 
if we assume that the difference between the two distribution is due to observational 
biases and the distribution was the same than the low mass stars, the plot of 
Figure \ref{fig8b} would imply an extremely high planet occurrence 
for the massive stars.

{\bf We shall therefore consider as an observational evidence  that 
planets around giants (more massive stars) are more massive than 
those around lower mass stars. If they were sharing a similar overall 
mass distribution, the frequency so far observed among giants would 
imply a very  high  planetary frequency around massive stars. }

As far as orbital  parameters are concerned, 
only long period planets are found around giants. This is quite expected since the stars 
have radii typically as large as 10 R$_{\odot}$ , therefore short period 
planets would be swallowed by the stars. It would be interesting to study 
the case of migration and planetary evolution in giants, in the presence of 
radius growth and mass losses, as occurring along the RGB and AGB phases. 
We note that the short period RV variations in giant stars may mask the detection of possible
short period planets that may still reside outside the photosphere of the star.
These would have orbital periods of many days, or similar to the periods for stellar
oscillations. Such variability if found by RV surveys may be dismissed as due to
stellar oscillations rather than a short period companion.

The distribution of eccentricity  is very similar to what is 
observed among main sequence 
stars and ranges from planets in nearly circular orbits to those in highly
eccentric orbits. 
It seems therefore that eccentricity is not affected by the 
stellar characteristics such as the mass. We note that there could be also here a hidden  bias 
in the eccentricity distribution: orbits with high eccentricity are much more 
difficult to be mimiced by other phenomena, like oscillations or stellar surface structure
so such RV variations are more  easily attributed to Keplerian motion. A nearly sinsoidal
RV variations can also arise from stellar surface structure and is thus more
suspicious. Additional tests (e.g. lack of photometric, Ca II H\&K, or line bisector
variations) are needed to confirm these as planetary companions. 

\section{ Interpretation}

It is natural to interpret the differences between giants' and main sequence stars 
starting from their three main differences: 
 giants are on average more massive, have larger convective zones and larger radii. 

\begin{figure}[h]
\begin{center}
 \includegraphics[width=3.0in]{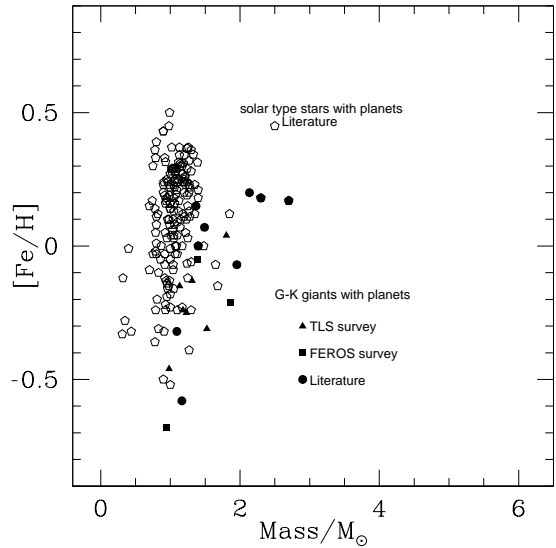} 
 \caption{Mass-Metalicity distribution of hosting - planets dwarfs and giants. 
 As expected, evolved stars are on average 
  more massive than main sequence stars.   }
   \label{fig9}
\end{center}
\end{figure}

Are giants hosting planets indeed more massive than their main sequence stars 
counterparts?  Given a volume-limited samples this would be  a natural result of stellar 
evolution,  and this difference is  confirmed  in Figure \ref{fig9}, which shows the 
mass-metallicity distribution for main sequence stars and giants hosting planets. 
Even if the giants we studied  are not really originating from very massive stars, their mass is,  on average, higher than that of  main sequence stars. 

As far as the tendency for planet-hosting giant stars not to be metal rich,
we favor the hypothesis 
that  the difference between main sequence and giants is due to pollution: 
the most external parts of main sequence atmospheres are polluted by 
engulfed debris and small planets which have a higher metal content. Because of the
relatively shallow convection zone of stars on the main sequence, these metals are
well mixed only in a small fraction of the star. 
During the giant phase the convective zone of the stars
greatly deepens and the more efficient mixing of the atmosphere dilutes the atmospheric 
abundance of metals.
This scenario has been debated at length in literature, 
and most analyses conclude against this hypothesis (Santos et al. 2004, 2006, Fischer and Valenti 
2005, Ecuvillon et al. 2006). We find that the counter arguments are not particularly strong, 
mostly because it is not very well known the real depth of the mixing zone in main sequence stars
and its variation with stellar mass (cfr. Vauclair 2006). By comparison
this effect is greatly enhanced 
(a factor $\sim$30 or more) as soon as
a star evolves and becomes a  red giant. 
A similar effect should happen at the very low mass end of the stellar mass distribution, 
where the stars become fully convective. 
If  highly convective M stars show a non-dependence on metallicity similar to the  
giants, as seems from the first analysis 
(Bonfils et al., these proceedings), then the evidence would be even stronger. 
The main observational argument against this interpretation stems in the subgiants analyzed
by Fischer and Valenti (2005), which do not observe any difference between 
subgiants in the blue and red part of the CMD. Stars on the two sides of the 
Hertzpung gap (i.e. before and after  the deep convective zones have developed)
 show no differences in metallicity.
We can offer no explanation for this  point.  
We already noticed as the metal distribution of subgiants shows a 
tail at very high metallicity which is not present in the giant samples analyzed so far, 
and we also note that  Murray et al. (2001) reached opposite conclusions than 
Fischer and Valenti. The issue is still open and we
believe that subgiants are worth a dedicated study. 

As far the trend that giant stars  might have a higher frequency of
more massive planets, stellar mass seems the most likely governing parameter, 
possibly because massive stars may develop  more massive proto-planetary disks.
Similar conclusions have been reached recently by two 
 papers (Lovis and Mayor 2007, Johnson et al. 2007b), which discussed stars in clusters and in
field low mass stars. It is also highly suggestive that while 
Paulson et al. (2004) 
did not find any planet around almost 100 low mass  stars surveyed in the Hyades, one planet has been 
found around one of the 2 single giants of this cluster (Sato et al. 2007). 
As far  to this topic, we can observe that all analyses so far agree with an enhanced frequency (and higher mass) of planets around more massive stars. 
We note also that a similar behavior is present in the samples of 
 Fischer and Valenti (2005) and Santos et al. 2006, but both authors 
conclude that this  dependence was uniquely the result of a selection bias, and that 
metallicity is the relevant parameter, because metallicity and stellar mass were 
strongly correlated in their samples. 
Since giants do not show a dependence of planet formation on metallicity 
and  widen the  range of stellar
masses observed, we propose that there may be a strong 
dependence of planet formation on stellar mass. Clearly, more studies are needed to disentangle
the possible effects of various stellar parameters (e.g. mass, abundance) on planet formation.

\section{ Giants and Planet Formation Theories}

Considering  the `classical'  planet  formation schemes:  core accretion 
(Ida \& Lin 2004) and disk instability (Boss 1997), 
we cannot conclude that the characteristics of the planets around giant stars 
favor either of the two schemes.   
It is on the other hand accepted that the metal dependence of massive planet
formation  in main sequence stars 
is considered a great success of the core accretion scenario (Ida \& Lin 2004). 

It is very possible that reality is  more complex than firstly envisaged; 
could it be that the  two mechanisms 
are both at work, and one is favored with respect to the other depending, for instance,
on the stellar mass? Something in this line has been proposed by Matsuo et al. (2007). 

On the other hand, a more complex core accretion scenario,
which involves the interaction with the 
stellar magnetosphere and the position of the dust evaporation  line for 
intermediate mass stars, in addition of the 
snow line considered in less massive stars, could also explain the 
observations (Lin , private communication). 

We do not know yet the answer, but clearly the 
extension of planet search to giants is  bringing new insides,
 which should be strongly pursued. 

\section{Next steps}
In this last section we would like to express our  vision for the 
continuation of this exciting research. 

Obviously, the results from other surveys are needed to confirm our results, and 
to better quantify them. Reaching smaller mass planets would also be 
essential, to determine the planet mass distribution  among intermediate mass stars. 
The results presented at this conference by other groups are extremely encouraging, 
and promise to bring a wealth of new results in a few years. 

We stress that a proper analysis of the hosting stars is essential, if we 
aim at obtaining results which can express more about physics.
The parent sample require to be analyzed as carefully as the planet hosting stars. 

Open clusters, where stars are supposed to share the same age and chemical 
compositions,   could bring direct new results. Similarly subgiants are worth 
to have a dedicated study. 
In principle it would be great to compare equivalent samples of field  giants and dwarfs, 
but this seems quite difficult, when all the variables (including age) are 
considered. 

Since ultimately we will aim at comparing stars in different evolutionary status, one important point
is to be sure that the analysis methods do not suffer of biases and peculiarities 
when applied at stars of different evolutionary status. Open clusters could be again 
privileged testbeds: analyzing stars in a well populated open cluster 
from main sequence to tip of RGB would be a crucial test to ensure that 
the results obtained do not suffer of any bias (see e.g. the analysis of 
Pasquini et al. 2004 of the open cluster IC4651).


\section{acknowledgements}
The observations have been collected at ESO and the 2m Alfred Jensch
Telescope of the Th\"uringer Landessternwarte Tautenburg.





\begin{thebibliography}{}
\bibitem[Biazzo et al. 2007a]{bi07a} Biazzo , K. et al. 2007a, AN 328, 938
\bibitem[biazzo 2007]{bia07} Biazzo, K., Pasquini, L. et al. 2007b, A\&A 475, 981
\bibitem[Boss 1997]{boss97} Boss, A. P. 1997, Science, 276, 1836
\bibitem[da Silva et al. 2006]{da} da Silva, L., Girardi, L.,  et al. 2006, A\&A 458, 603
\bibitem[D\"ollinger et al. 2007]{mic1} D\"ollinger. M.P., et al. 2007a: A\&A, 472, 649
\bibitem[D\"ollinger et al. 2008a]{mic2} D\"ollinger. M.P., et al. 2007b: A\&A, in preparation
\bibitem[D\"ollinger et al. 2008b]{mic3} D\"ollinger. M.P., et al. 2007c: A\&A, in preparation
\bibitem[Ecuvillon et al. 2006]{ec06} Ecuvillon, A., Israelian, G.,  et al. 2006, A\&A, 449, 809
\bibitem[Fischer \&\ Valenti 2005]{fischerValenti05} Fischer, D., \& Valenti, J. 2005, ApJ, 622, 1102 
\bibitem[Frandsen et al. 2002]{fra02}
Frandsen, S., Carrier, F., Aerts, C. et al. 
 2002, A\&A, 394, L5.
\bibitem[Gonzalez, G. 1997]{gonz97} Gonzalez, G. 1997, MNRAS, 285, 403
\bibitem[Gonzalez, G. 1998]{gonz98} Gonzalez, G. 1998, A\&A, 334, 221 
\bibitem[Gonzalez et al 2001]{gonz01} Gonzalez, G., Laws, C. et al. 2001, AJ, 121, 432 
\bibitem[Gray \& Johnes 1991]{gj91} Gray, D. Johnes, H. 1991 PASP 103, 409
\bibitem[Hatzes \&\ Cochran1993]{hatzes93} Hatzes, A.P., Cochran, W.D. 1993, ApJ, 413, 339
\bibitem[Hatzes \&\ Cochran1994]{hatzes94} Hatzes, A.P., Cochran, W.D.
 1994, ApJ, 422, 366
\bibitem[Hekker \& Melendez 2007]{hm07} Hekker, Melendez 2007 astro-ph 07091145
\bibitem[Ida \& Lin 2004]{idaLin04} Ida, S., \& Lin, D. N. C. 2004, ApJ, 616, 567
\bibitem[Johnson et al. 2007a]{john07a} Johnson, J. et al. 2007a ApJ 665, 785
\bibitem[Johnson et al. 2007a]{john07a} Johnson, J. et al. 2007b ArXiv 0707.0518
\bibitem[]{jor05}  Jo{$\!\!\!/$}rgensen, B.R., Lindegren, L. 2005, A\&A, 436, 127
\bibitem[laad]{la97}  Laughlin, G., \& Adams, F. C. 1997, ApJ, 491, L51  
\bibitem[Lovis \& Mayor 07]{lo07} Lovis, C., Mayor, M. 2007 A\&A 472, 657
\bibitem[Matsuo et al. 2007]{ma07} Matsuo, T., et al. 2007, ApJ, in press (astro-ph 0703237)
\bibitem[Mayor \&\ Queloz ]{mayorqueloz95} Mayor, M., \& Queloz, D. 1995, Nature, 378, 355 
\bibitem[Murray et al. 2001]{murray01} Murray N., Chaboyer B.   2001, ApJ 555, 801
\bibitem[Pasquini et al. (2004)]{pas04} Pasquini, L., Randich, S. et al. 2004, A\&A 424, 951
\bibitem[Pasquini et al. (2007)]{pas07} Pasquini, L., D\"ollinger, M. Weiss, A. et al. 2007, A\&A 473, 979
\bibitem[Paulson et al. (2004)]{pau04} Paulson, D.B., Cochran, B., Hatzes, A.P. 2004, AJ 127, 3579
\bibitem[Pollack et al. 2006]{oollack06}  Pollack, J. B., Hubickyj, O.,  et al. 1996, Icarus, 124, 62
\bibitem[Sadakane et al. 2005]{sad05} Sadakane, K., Ohnishi, T., Ohkubo, M., Takeda, Y. 2005, PASJ 57, 127 
\bibitem[Santos et al. 2000]{santos00} Santos, N. C., Israelian, G.,  Mayor, M. 2000, A\&A, 363, 228
\bibitem[Santos et al. 2001]{santos01} Santos, N. C., Israelian, G., Mayor, M. 2001,  A\&A, 373, 1019
\bibitem[Santos et al. 2003]{santos03}Santos, N. C., Israelian, et al. 2003, A\&A, 398, 363  
\bibitem[Santos et al. 2004]{santos04} Santos, N.C., Israelian, G., Mayor, M. 2004, A\&A, 415, 1153
\bibitem[Santos et al. 2004]{santos05}  Santos, N. C., Israelian, G., et al. 2005, A\&A, 437, 1127
\bibitem[Sato2003]{sato03} Sato, B., Ando, H., Kambe, E. 2003, ApJ, 597, L157
\bibitem[Sato et al. (2007)]{sat07} Sato, B., Izumiura, H., et al. 2007, ApJ, 661, 527 
\bibitem[Schuler et al. (2005)]{sch05} Schuler, S., Kim, J.H. et al. 2005, ApJ, 632, L131
\bibitem[Setiawan et al. 2003a]{setiawan03a} Setiawan, J., Hatzes, A.P., et al. 2003a, A\&A, 398, L19
\bibitem[Setiawan 2003b]{setiawan03b} Setiawan, J., Pasquini, L., et al. 2003b, A\&A, 397, 1151
\bibitem[Setiawan et al. (2004)]{set04} Setiawan, J., Pasquini, L., et al. 2004, A\&A 421, 241
\bibitem[Setiawan et al. 2005]{setiawan05} Setiawan, J., Rodman, J.,  et al. 2005, A\&A, 437, L31     
\bibitem[Vauclair 2004]{vauclair04} Vauclair S.  2004, ApJ 605, 874
\bibitem[Walker et al. (1989)]{wal89} Walker, G.A.H., Yang, S., Campbell, B., Irwin, A.W. 1989, ApJ, 343, L21.
\end{thebibliography}
\end{document}